# Impact of Mobility On QoS of Mobile WiMax Network With CBR Application

**Kranti Bala**[*], **Kiran Ahuja**

Department of Electronics and Communication Engineering,
DAV Institute of Engg. & Technology, Jalandhar, Punjab, India.
*kalpana_31282@yahoo.com, askahuja2002@yahoo.co.in

*Abstract*

The issue of mobility is important in wireless network because internet connectivity can only be effective if it's available during the movement of node. To enhance mobility, wireless access systems are designed such as IEEE 802.16e to operate on the move without any disruption of services. In this paper we are analyzing the impact of mobility on the QoS parameters (Throughput, Average Jitter and Average end to end Delay) of a mobile WiMAX network (IEEE 802.16e) with CBR application.

## 1. Introduction

Providing diverse broadband services every time to mobile subscribers will be a major challenge for the telecommunication community [1]. Mobile WiMAX (Worldwide Interoperability for Microwave Access) introduces the most significant new feature i.e. mobility to support for handovers; this can be considered as a basic requirement for mobile communication system. In order to achieve the high data rate in wireless services such as VoIP and IPTV [2], Mobile WiMAX based on the IEEE 802.16e standard is developed as broadband wireless solution to the wired backhaul. Mobile WiMAX covers up to thirty mile radius and data rates between 15 Mbps to 75 Mbps theoretically. IEEE 802.16e supports four types of mobility i.e. nomadic, portable, simple mobility and full mobility [3]. Nomadic mobility means user is allowed to take a fixed subscriber station and reconnect from a different point of attachment. Portable mobility means nomadic access is provided to a portable device, such as a PC card, with expectation of a best-effort handover. In simple mobility the subscriber may move at speeds up to 60 km/h (Kilometer per Hour) with brief interruptions (less than 1 sec) during handover and Full mobility supports up to 120 km/h speed and seamless handover (less than 50 ms latency and < 1% packet loss). The enhanced IEEE 802.16e system has the capability to fulfill the requirements regarding the mobility management of future telecommunication systems. Major element introducing complexity in mobility is the need of handovers. A typical geographical area cannot be covered by one base station, necessitating the design of cellular networks [4]. Handover operation is the process when a mobile user goes from one cell to another without interruption of the ongoing session (whether phones call, data session or other). The handover can be due to movement of mobile subscriber or due to change in radio channel condition or due to cell capacity constraints.

The paper is organized as follows: Section II briefly outlines the related work. Section III describes the scenario under consideration. Section IV show results and discussions and then it is concluded in section V

## 2. Related Work

Researchers have done lot of work in the field of WiMAX (IEEE802.16) and Mobile WiMAX (IEEE802.16e). A standard that specifies the air interface of fixed broadband wireless access (BWA) systems supporting multimedia services was given in [1]. The medium access control layer (MAC) supports a primarily point-to-multipoint architecture, with an optional mesh topology. Then enhancements to IEEE Std 802.16 2004 was introduced in 2006 to support subscriber stations moving at vehicular speed [2] and thereby specified a system for combined fixed and mobile broadband wireless access. An overview of Mobile WiMAX and the performance for the basic minimal configuration based on the WiMAX Forum Release-1 system profiles was given in [3].

Members of WiMAX forum provided a comparison with contemporary cellular alternatives in [4]. Authors of [5] particularly looked at the advanced features introduced by the IEEE802.16e standard for mobility in WiMAX systems in order to prove that the enhanced IEEE 802.16e system has the capability to fulfill the requirements regarding the mobility management of future telecommunication systems. In [6], authors provided an overview of the state-of-the-art mobile WiMAX technology and its development. They expected that future work will be focused on the mobility aspect and interoperability of mobile WiMAX with other wireless technologies. In [7] researchers analyzed the performance of a mobile WiMAX system for various link speeds in an urban microcell and simulated results were compared with measured drive test results from a carrier-class WiMAX base station. Authors in [8] analyzed both the competitive and cooperative relationships between WiMAX, WLAN and 3G from various aspects, such as technical standards, current status, future trend and marketing orientation, etc. Thus a brand-new future for mobile communication is highly expectable. Numbers of mobility improvement algorithms were presented earlier, but the authors of [9] presented a mobility improvement handover algorithm with reduced scan time implementation for Mobile WiMAX. So this creates interest for analysis of the impact of mobility in WiMAX network. In next section we develop a system for the same.

## 3. System Description

We develop a scenario using Qualnet 5.0 to analyze the impact of mobility on QoS of WiMAX (IEEE 802.16e) network with CBR (Constant Bit Rate) application between two mobile stations. CBR is data traffic that keeps the bit rate same throughout the process. Figure 1 shows the general view of scenario of mobile WiMAX (IEEE 802.16e) network.

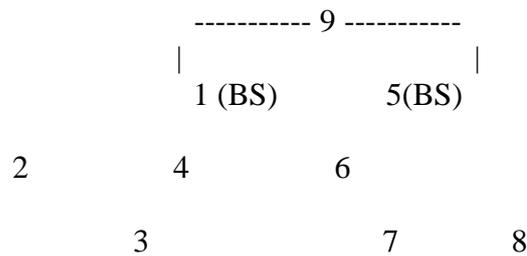

Figure 1 General view of scenario of Mobile WiMAX (IEEE 802.16e) network

In this Scenario we are considering two WiMAX networks 192.0.0.0 and 192.0.1.0. WiMAX network 192.0.0.0 has four nodes (1 to 4) with 1 as base station (BS) and the rest are mobile stations. WiMAX network 192.0.1.0 has four nodes (5 to 8) with 5 as base station (BS) and the rest are mobile stations. Node 9 is also a mobile station of another WiMAX network. Both nodes 1 and 5 are connected to node 9 via wired point to point links. The two base stations are operating on different wireless channels. Node 1 is operating on channel 0 and Node 5 is operating on channel 1.

We consider random waypoint mobility model in which mobile node is allowed to move randomly in a specified area. Node 3 is originally close to BS node 1 and it is register with BS node 1. Node 3 will move from left to right and CBR application is used between nodes 3 and 9. As per scenario, node 3 is server, receiving packets and node 9 is client, sending packets. Figure 2 represents the scenario to analyze the impact of mobility on QoS parameters of mobile WiMAX (IEEE 802.16e) network with CBR application between two mobile stations using Qualnet 5.0

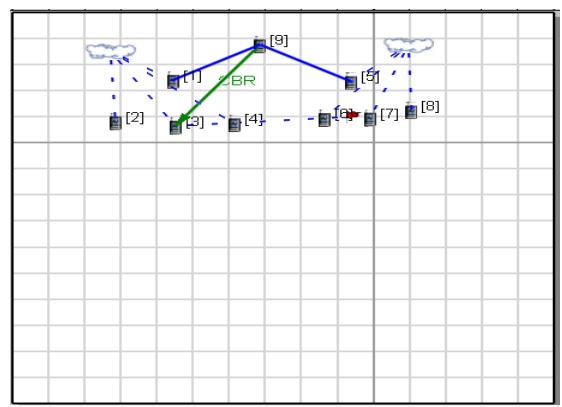

Figure 2 Scenario to analyze mobility of IEEE 802.16e using Qualnet 5.0

As node 3 moves right, it will perform handover with BS node 5. Connectivity on the move can only be ensured with fast and efficient handovers. The handover process is defined as the set of procedures and decisions that enables an MS to migrate from the air interface of one BS to that of another. As per the above said scenario, following results have been taken which shows the impact of mobility on QoS of mobile WiMAX (IEEE 802.16e) network with CBR application.

## 4. Results And Discussions

By simulation of the scenario developed for analysis of mobility impact on QoS parameters of mobile WiMAX (IEEE 802.16e) network with CBR application, we observe the following results with QoS parameters (Throughput, Average Jitter, Average end to end delay).

*Throughput:* It refers to how much data can be transferred from one location to another in a given amount of time. It is measured in bps (bit per second). Client throughput is 4096 bps, number of bytes sent by client are 97280 bps and number of packets sent by client are 190. All these parameters of client will remain constant.

*Average Jitter:* - It is a variation or dislocation in the pulses of a digital transmission; it may be in the form of irregular pulses.

*Average End-to-end delay:* - It refers to the time taken for a packet to be transmitted across a network from source to destination.

$$d_{end-end} = N[\ d_{trans} + d_{prop} + d_{proc}] \qquad (1)$$

Where $d_{end-end}$ is end-to-end delay, $d_{trans}$ is transmission delay, $d_{prop}$ is propagation delay, $d_{proc}$ is processing delay and N is number of links.

To analyze the impact of mobility speed is compared with throughput, average jitter and average end to end delay respectively.

*Speed Vs. Throughput:* - In this scenario we consider CBR application between 9 and 3 mobile nodes, so data or packets are transferred from node 9 to node 3. Server throughput will vary according to data or packets received by node 3. The variation of throughput with respect to change in speed is shown in figure 3

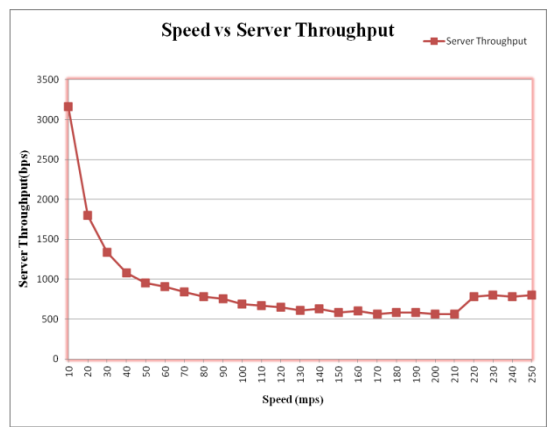

Figure 3: Speed vs. Server throughput to analyze the impact of mobility

According to figure 3, when speed is 10 mps (meter per sec) throughput is 3160 bps, with the increase in speed throughput is decreasing. At speed of 210 mps throughput is minimum i.e. 562 bps. Node 3 is close to BS 1 and is registered with it. As node 3 moves from left to right, it will perform handover and will go far away from its original registered BS and during that time there is loss of data (bytes and packets) sent by node 9 to node 3. After that as speed increases the throughput will start increasing again because we consider random waypoint mobility model, therefore when node 3 again comes closer to its original registered BS 1 data loss will decrease.

***Speed Vs. Average Jitter: -*** Jitter is basically due to connection timeouts, connection time lags, data traffic congestion, and interference. Variation of average jitter with respect to speed is shown in figure 4.

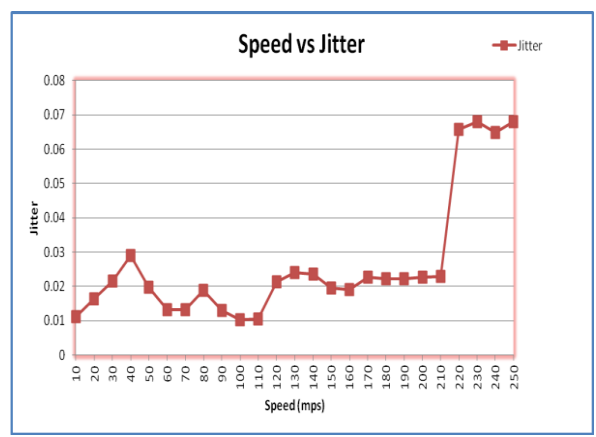

Figure 4:  Speed vs. Average jitter to analyze the impact of mobility

According to figure 4, jitter is very less initially i.e.0.01129s when node 3 is moving with less speed of 10mps; with increase in speed jitter will also increase i.e.  0.06792s at 250 mps speed. The reason behind it is that data is split up into manageable 'packets' with headers and footers that indicate the correct order of the data packets or whole signal is broken down into chunks of data which is transmitted to a receiving unit for assembly. If jitter occurs, synchronization becomes a problem and the receiving unit finds it difficult to correctly assemble the incoming data stream**.** Therefore at high speed jitter is more and throughput is less.

***Speed Vs. Average End To End Delay: -*** In this scenario, node 9 is sending packets towards node 3 because CBR application is connected between these two nodes. During transmission, delay is introduced and average end to end delay changes with the change in speed. Variation of average end to end delay with respect to speed is shown in figure 5.

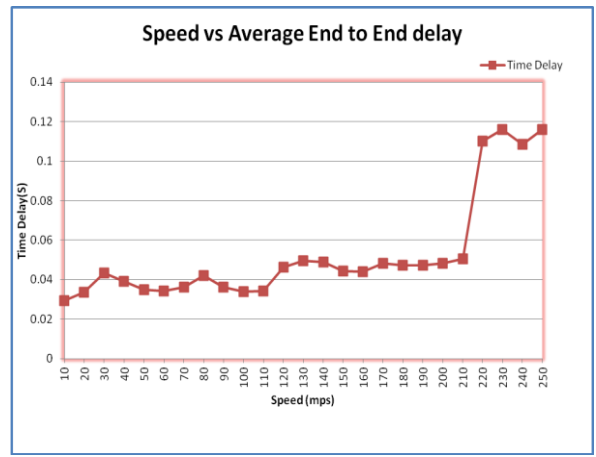

Figure 5: Speed Vs. Average End To End Delay To Analyze The Impact Of Mobility

According to figure 5, average end to end delay is very less when speed is less i.e. 0.02926 s at the speed of 10 mps but as speed increases average end to end delay will decrease i.e. 0.11596 at the speed of 250 mps. Average end to end delay increases or decreases according to movement of node 3 as the mobility model is Random waypoint. All results observed using the Scenario to analyze impact of mobility on QoS parameters are tabulated in table1.

Table 1:  Impact of mobility (speed) on QoS parameters of mobile WiMAX network with CBR application

| **Speed (mps)** | **10** | **50** | **100** | **150** | **200** | **250** |
|---|---|---|---|---|---|---|
| **Server throughput (bps)** | 3160 | 950 | 691 | 584 | 562 | 800 |
| **Average jitter(s)** | 0.011 | 0.01 | 0.01 | 0.01 | 0.02 | 0.06 |
| **Average end to end delay (s)** | 0.029 | 0.03 | 0.03 | 0.04 | 0.04 | 0.11 |

**5. Conclusion**

It is concluded that as mobile node moves away from registered Base Station at high speed; throughput will decrease because some of the data is lost due to handover. Jitter and end to end time delay also vary with the variation in speed; initially these were very less. When speed is 10 mps jitter is 0.01129s and average end to end delay is 0.02926 and when speed is 250 mps then jitter increases to 0.06792s and average end to end delay increase to 0.11596s. As handover takes place, then due to connection timeouts, connection time lags, data traffic congestion, and interference jitter occurs and because of this average end to end delay increases, but when mobile node again come to its registered base station both jitter and average end to end delay will decrease although. As we are using random waypoint mobility model, throughput will increase further when mobile node again come under its original registered base station.